\documentclass[12pt,thmsa]{article}
\usepackage{sw20lart}

\input{tcilatex}
\begin{document}

\title{On a recent experiment seemingly violating quantum predictions}
\author{Emilio Santos \\
Departmento de F\'{i}sica, Universidad de Cantabria. Santander. Spain}
\maketitle

\begin{abstract}
A disagreement of the empirical results with quantum mechanical predictions
is pointed out in the experiment by M. Giustina et al.
\end{abstract}

A recent experiment has violated for the first time a Bell inequality with
photons without the fair sampling assumption\cite{1}. The purpose of this
note is to point out that one of the correlations measured in the experiment
does not agree with the quantum mechanical prediction.

In the experiment entangled (not maximally) photon pairs are produced by
parametric down conversion in such a way that the quantum state of the pair
may be represented by 
\begin{equation}
\mid \Psi \rangle =\frac{1}{\sqrt{1+r^{2}}}\left( \mid HV\rangle +r\mid
VH\rangle \right) ,  \label{1}
\end{equation}
with $r=0.297$ and $H(V)$ denotes horizontal (vertical) polarization of
Alice's and Bob's photons. The quantum prediction for the probability of a
coincidence count with the measuring devices placed at angles $\alpha $ and $%
\beta $ is 
\begin{equation}
p_{AB}\left( \alpha ,\beta \right) =\frac{1}{1+r^{2}}\left( \cos \alpha \sin
\beta +r\sin \alpha \cos \beta \right) ^{2},  \label{2}
\end{equation}
and the probabilities of single counts are 
\begin{eqnarray}
p_{A}\left( \alpha \right) &=&\frac{1}{1+r^{2}}\left( \cos ^{2}\alpha
+r^{2}\sin ^{2}\alpha \right) ,  \nonumber \\
p_{B}\left( \beta \right) &=&\frac{1}{1+r^{2}}\left( \sin ^{2}\beta
+r^{2}\cos ^{2}\beta \right) .  \label{3}
\end{eqnarray}

In the experiment four correlations, $C\left( \alpha _{i},\beta _{j}\right)
, $ $i,j=1,2$, were measured for a total of 300 seconds per setting at each
of the four settings described by the angles $\alpha _{1}=85.6{{}^{o}},$ $%
\alpha _{2}=118.0{{}^{o}},$ $\beta _{1}=-5.4{{}^{o}},$ $\beta _{2}=25.9{%
{}^{o}.}$ Also two single counts, $S\left( \alpha _{1}\right) ,S\left( \beta
_{2}\right) $ were measured. The singles and coincidence counts obtained
appear below in the first raw of Table 1, taken from the published paper\cite
{1}. As a result of these data the authors report a J-value corresponding to
a violation of the measured Bell inequality by $69-\sigma .$ The J-value is
defined by 
\begin{eqnarray}
J &=&C\left( \alpha _{1},\beta _{1}\right) +C\left( \alpha _{1},\beta
_{2}\right) +C\left( \alpha _{2},\beta _{1}\right) -C\left( \alpha
_{2},\beta _{2}\right) -S\left( \alpha _{1}\right)  \nonumber \\
-S\left( \beta _{1}\right) &=&-126715,  \label{5}
\end{eqnarray}
and it should be non-negative for any local hidden variables theory.

For comparison the single and coincidence counts predicted from eqs.$\left( 
\ref{2}\right) $ and $\left( \ref{1}\right) $ are given in the second raw of
Table 1. They are calculated using the estimated number of produced pairs
per setting $24.2\cdot 10^{6},$ and the arm efficiencies $\eta
_{A}=73.77\%,\eta _{B}=78.59\%$\cite{1}$.$

\textbf{Table 1}. Comparison between the results of the experiment and the
quantum prediction. Numbers in the first (second) raw correspond to the data
of the experiment (quantum prediction) ($\times 1000)$.

$
\begin{array}{lllllll}
& S\left( \alpha _{1}\right) & S\left( \beta _{1}\right) & C\left( \alpha
_{1},\beta _{1}\right) & C\left( \alpha _{1},\beta _{2}\right) & C\left(
\alpha _{2},\beta _{1}\right) & C\left( \alpha _{2},\beta _{2}\right) \\ 
Exper. & 1523 & 1694 & 1069 & 1153 & 1191 & 69.79 \\ 
QM & 1535 & 1683 & 1066 & 1160 & 1201 & 12.25
\end{array}
$

The disagreement between the empirical data and the quantum predictions in
the former five columns might be explained by experimental errors. Indeed
they are only a few times larger than the expected statistical
uncertainties. In contrast there is a dramatic difference in the latter
correlation $C\left( \alpha _{2},\beta _{2}\right) ,$ where the empirical
result is more than four times the quantum prediction.

I shall point out that ``nonocurrence of coincidences'' for some
combinations of angles is hardly compatible with local realism (as stressed
long ago by Jaynes\cite{2}.) Therefore it is worth measuring correlations
for angles where the quantum prediction is zero. In the commented experiment%
\cite{1} the prediction for the correlation $C\left( \alpha _{2},\beta
_{2}\right) ,$ although not strictly zero, is one about hundred times
smaller than any other of the three measured correlations. Of course Bell
inequalities provide better tests of local realism, but the anomaly here
discussed may be an indication that a loophole-free Bell test, if possible,
might not give results refuting local realism. Furthermore the commented
experiment is not the first one exhibiting the anomaly\cite{3}. For these
reasons a careful investigation of the matter is worth while.

\end{document}